%% file: CA-V3.1-2023.tex
\documentclass[letterpaper,twocolumn,prl,aps,superscriptaddress,floatfix,longbibliography]{revtex4-2}
\usepackage[latin1]{inputenc}
\usepackage[T1]{fontenc}
\usepackage{bm}
\usepackage{amssymb,amsbsy,amsmath}
\usepackage{graphicx}
\usepackage{verbatim}
\usepackage{color}
\usepackage{pifont}
\usepackage{hyperref}
\usepackage{xcolor}
\usepackage{verbatim}
\usepackage{algorithm}
\usepackage{algorithmicx}
\usepackage{algpseudocode}
\usepackage{float}
\usepackage{algorithm,algpseudocode,float}
\usepackage{lipsum}



\hypersetup{
	colorlinks=true,       		
	linkcolor=orange,          	
	citecolor=blue,             
	urlcolor=blue,          
}
\usepackage{braket}

\begin{document}

	\title{Realization of algorithmic identification of cause and effect in quantum correlations}

    \author{Zhao-An~Wang}
    \author{Yu~Meng}
    \email{mengyu23@mail.ustc.edu.cn}
    \author{Zheng-Hao~Liu}
    \altaffiliation[Present address: ]{Center for Macroscopic Quantum States (bigQ), Department of Physics, Technical University of Denmark, Kongens Lyngby, Denmark}
    \affiliation{CAS Key Laboratory of Quantum Information, University of Science and Technology of China, Hefei, 230026, China}
    \affiliation{CAS Center For Excellence in Quantum Information and Quantum Physics, University of Science and Technology of China, Hefei, 230026, China}
    \author{Yi-Tao~Wang}
    \affiliation{CAS Key Laboratory of Quantum Information, University of Science and Technology of China, Hefei, 230026, China}
    \affiliation{CAS Center For Excellence in Quantum Information and Quantum Physics, University of Science and Technology of China, Hefei, 230026, China}
    \author{Shang~Yu}
    \affiliation{Department of Physics, Imperial College London, Prince Consort Rd, London SW7 2AZ, UK}
    \affiliation{CAS Key Laboratory of Quantum Information, University of Science and Technology of China, Hefei, 230026, China}
    \affiliation{Research Center for Quantum Sensing, Zhejiang Lab, Hangzhou, 310000, People's Republic China}
    \author{Wei~Liu}
    \author{Zhi-Peng~Li}
    \author{Yuan-Ze~Yang}
    \author{Nai-Jie~Guo}
    \author{Xiao-Dong~Zeng}
    \affiliation{CAS Key Laboratory of Quantum Information, University of Science and Technology of China, Hefei, 230026, China}
    \affiliation{CAS Center For Excellence in Quantum Information and Quantum Physics, University of Science and Technology of China, Hefei, 230026, China}
    \author{Jian-Shun~Tang}
    \email{tjs@ustc.edu.cn}
    \affiliation{CAS Key Laboratory of Quantum Information, University of Science and Technology of China, Hefei, 230026, China}
    \affiliation{CAS Center For Excellence in Quantum Information and Quantum Physics, University of Science and Technology of China, Hefei, 230026, China}
    \affiliation{Hefei National Laboratory, University of Science and Technology of China, Hefei 230088, China}
    \author{Chuan-Feng~Li}
    \email{cfli@ustc.edu.cn}
    \affiliation{CAS Key Laboratory of Quantum Information, University of Science and Technology of China, Hefei, 230026, China}
    \affiliation{CAS Center For Excellence in Quantum Information and Quantum Physics, University of Science and Technology of China, Hefei, 230026, China}
    \affiliation{Hefei National Laboratory, University of Science and Technology of China, Hefei 230088, China}
    \author{Guang-Can~Guo}
    \affiliation{CAS Key Laboratory of Quantum Information, University of Science and Technology of China, Hefei, 230026, China}
    \affiliation{CAS Center For Excellence in Quantum Information and Quantum Physics, University of Science and Technology of China, Hefei, 230026, China}
    \affiliation{Hefei National Laboratory, University of Science and Technology of China, Hefei 230088, China}

	\date{\today}
	
	\renewcommand{\figurename}{Fig.}
	
	\newcommand{\Todos}[1]{\textcolor{red}{#1}}
			
\begin{abstract}

Causal inference revealing causal dependencies between variables from empirical data has found applications in multiple sub-fields of scientific research. A quantum perspective of correlations holds the promise of overcoming the limitation by Reichenbach's principle and enabling causal inference with only the observational data. However,
it is still not clear how quantum causal inference can provide operational advantages in general cases. 
Here, we have devised a photonic setup and experimentally realized an algorithm capable of identifying any two-qubit statistical correlations generated by the two basic causal structures under an observational scenario, thus revealing a universal quantum advantage in causal inference over its classical counterpart.
We further demonstrate the explainability and stability of our causal discovery method which is widely sought in data processing algorithms. 
Employing a fully observational approach, our result paves the way for studying quantum causality in general settings. 
\end{abstract}		

\maketitle
\textit{Introduction.}---Causal identification has attracted significant research interest in big data science. Unlike the human brain capable of thinking causally to some extent, many current artificial intelligences based on black-box models excel at finding correlations but struggle to identify causal dependencies. 
The obtained models consequently come with two drawbacks. Firstly, they are hard to interpret and can have unexpected behaviors: image classifiers for covid diagnosis can focus instead on confounders\,\cite{DeGrave21}; secondly, they could become unstable and get confused by small changes in the input data: self-driving cars working well on the Europe continent could become confused on the British islands\,\cite{Savage23}. 
The ability to distinguish between causes and effects thus holds the promise of enhancing the explainability and stability of the acquired knowledge\,\cite{cui2022stable}. 


The obstacle for machine learning-based methods to understand causality originates from a renowned limitation: Reichenbach's common cause principle (RCCP)\,\cite{Reichenbach} postulates that a latent variable co-influencing two events can always produce the same correlation as when the two events are linked by a direct causal link.
The implication of RCCP for causal identification is that distinguishing between a common cause (CC) and a direct cause (DC) is not possible using observational data generated by classical objects. 
Given that quantum correlations have drastically different behaviors\,\cite{Bell64, KS67, LG85} from their classical counterparts, it is natural to ponder to what extent a quantum perspective of correlations and causations overcomes the limitation from RCCP. The question has motivated a number of works exploring the quantum advantages in identifying causal relationships\,\cite{Cavalcanti2014, Henson2014,  Fitzsimons15, Ried15, Chaves2015, Costa2016, Allen17, Chaves18, mdhu18, barrett2019}. 

However, none of these works has enabled unconditional discrimination between the two most basic causal structures, i.e., to differentiate DC from CC with only observational quantum correlations. Here, we directly address the quantum analogy of the two-point causal identification problem. Our objective is to decide whether the causal structure of a two-point qubit correlation is induced by the same particle going through a unitary quantum channel which amounts to a DC, or two (possibly entangled) particles being successively measured which constitutes a CC. 


We have realized an optical setup that allows the switching between the two different mechanisms with a photonic controlled-\textsf{SWAP} gate. Using the setup, we experimentally demonstrate that a few judiciously chosen measurements suffice to distinguish any two-point qubit DC from CC in a geometric way. The measurement strategy was initially given in Reference\,\cite{cgzhang20}; here, we further develop the strategy into a systematic algorithm that is very robust to noise. Our results show the quantum advantage in two-point causal inference is universal, explainable, and stable and thus could be highly interesting to quantum machine learning.


\begin{figure}[t]
    \centering
    \includegraphics[width=0.48\textwidth]{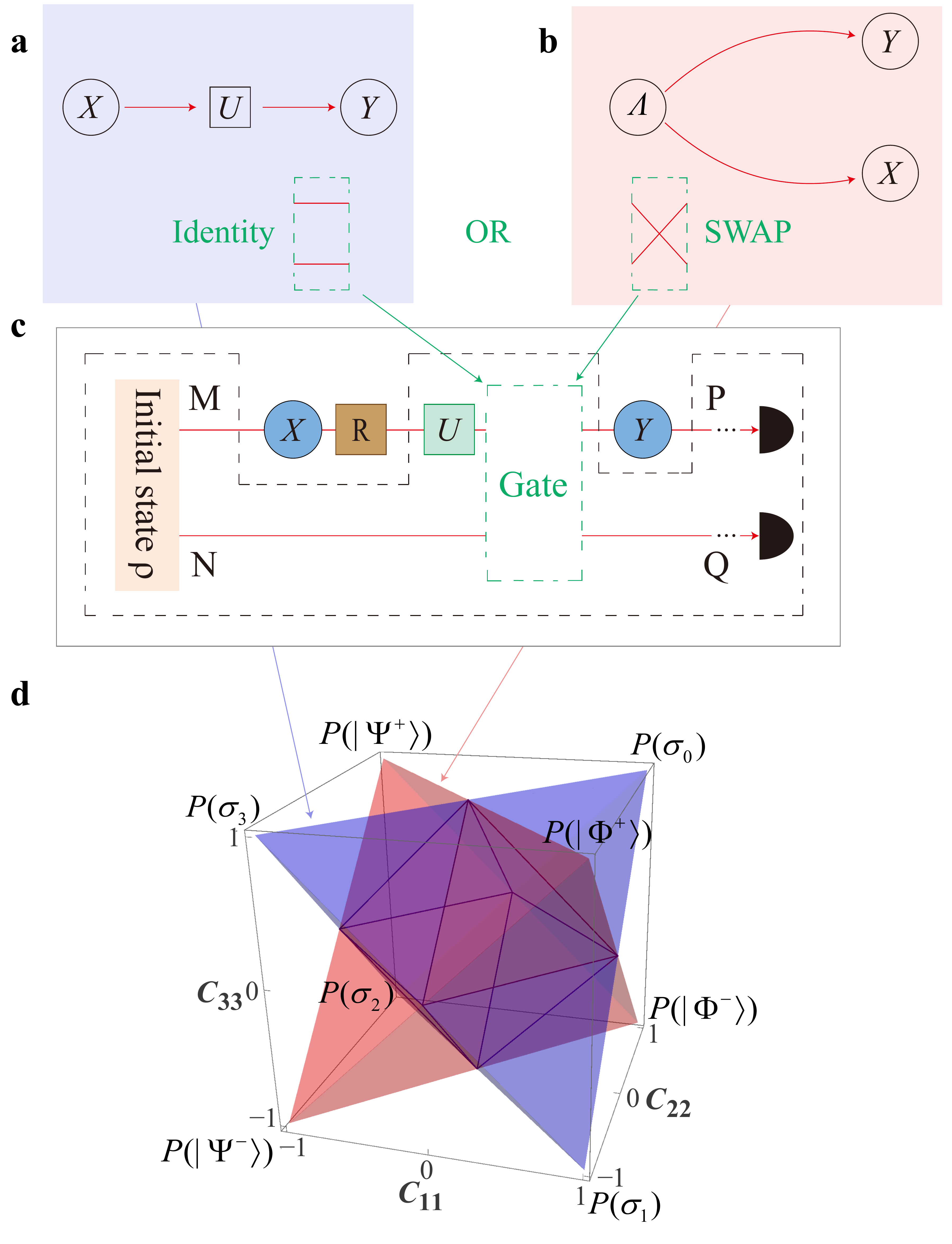}
    \caption{\textbf{Perspectives of two-point quantum causality.} 
    \textbf{a, b}: Directed acyclic graph of direct cause (DC) and common cause (CC).
    \textbf{c}: Quantum comb representation of causal structures. DC and CC can be switched by selecting the quantum gates in the green dashed box to be identity or \textsf{SWAP}. 
    \textbf{d}: Geometric description of two-qubit causal structures. The two-point correlations are characterized by pairs of identical Pauli measurements at the end of \textbf{c}. The outcomes are represented by the three-dimensional coordinate of a point $\mathbf{P}$. The blue and red tetrahedrons are the full set of points derived from DC and CC, respectively. The vertices of the DC and CC tetrahedrons are achieved using Pauli unitaries and Bell states.}
    \label{fig_scheme}
\end{figure}


\textit{Quantum causality with geometry.}---We start by introducing the quantum versions of the two most fundamental causal structures, DC and CC. In classical causal theory, a DC indicates an earlier variable directly affects a later variable, and a CC signifies a confounding factor co-influencing the two variables. 
In quantum theory, the notion of variables is replaced by quantum states. As shown in Fig.\,\ref{fig_scheme}a and b, a quantum channel linking its input and output states can act as a quantum version of DC, while a bipartite system comprising two subsystems gives rise to a quantum CC. Throughout the present work, we focus on two-qubit entanglement and unitary channels.


The causal structures of quantum systems cannot be trivially revealed by their sequence of emergence: a memory system could put two CC-linked subsystems into a timelike separation. In Fig.\,\ref{fig_scheme}c, we plotted a simple way to switch between DC and CC in the quantum comb representation\,\cite{Chiribella07, Chiribella21} where the observer has no access to the quantum process inside the dashed box. When the green-colored gate is chosen to preserve or swap the two subsystems, the quantum systems at $X$ and $Y$ will conform to a causal structure of DC induced by a unitary channel with its Kraus operator being $U$, or CC from the entangled initial state $\rho$, respectively.
As such, to infer the causal structure, an observer must rely on quantum measurements to extract further information. The measurements can have multiple settings and be destructive as long as they do not introduce signaling toward the other measurement. Therefore, a procedure
$R$ (cf.\ Fig.\,\ref{fig_scheme}c) is necessary to recover the post-measurement state according to the L\"uders' rule\,\cite{Luders}. This guarantees a preceding measurement will not change the marginal distribution of the next measurement and the causal inference scheme is strictly observational.

With the above preliminaries, we can recover the affirmation in Reference\,\cite{Ried15} that observational quantum causal identification is sometimes possible even with fixed combinations of measurements. The idea can be clearly illustrated in a geometric way: denote $i, j\in\{1, 2, 3\}$ to be the measurement settings at $X$ and $Y$ and the three indices mean the observables should be the three Pauli matrices $\sigma_1, \sigma_2$, and $\sigma_3$, respectively. 
Further, we define the following conditional probability:
\begin{align}
    C_{ij} := p(x=y|i,j)-p(x \neq y| i, j),
\end{align}
where $x,y$ are the outcomes of the measurements at $X$ and $Y$. The two-point correlations of the same-setting measurements can then be to a three-dimensional vector in the coordinate space:
 \begin{align}
    \mathbf{P} = \left(
    C_{11} \; \; C_{22} \; \; C_{33}
    \right)^{\rm T}.
\end{align}
We use $\mathbf{P}(U)$ and $\mathbf{P}(\rho)$ to differentiate the correlations generated from DC and CC and plot the sets of all possible $\mathbf{P}(U)$ and $\mathbf{P}(\rho)$ in Fig.\,\ref{fig_scheme}d as the blue and red polytopes, which we refer to as DC- and CC-tetrahedra. 
The tetrahedral shape follows from that any two-point qubit correlation can be expressed as the convex combination of the four extremal points at the corresponding tetrahedron's vertices\,\cite{mdhu18}. 

\begin{figure*}[t]
    \centering
    \includegraphics[width=.99\textwidth]{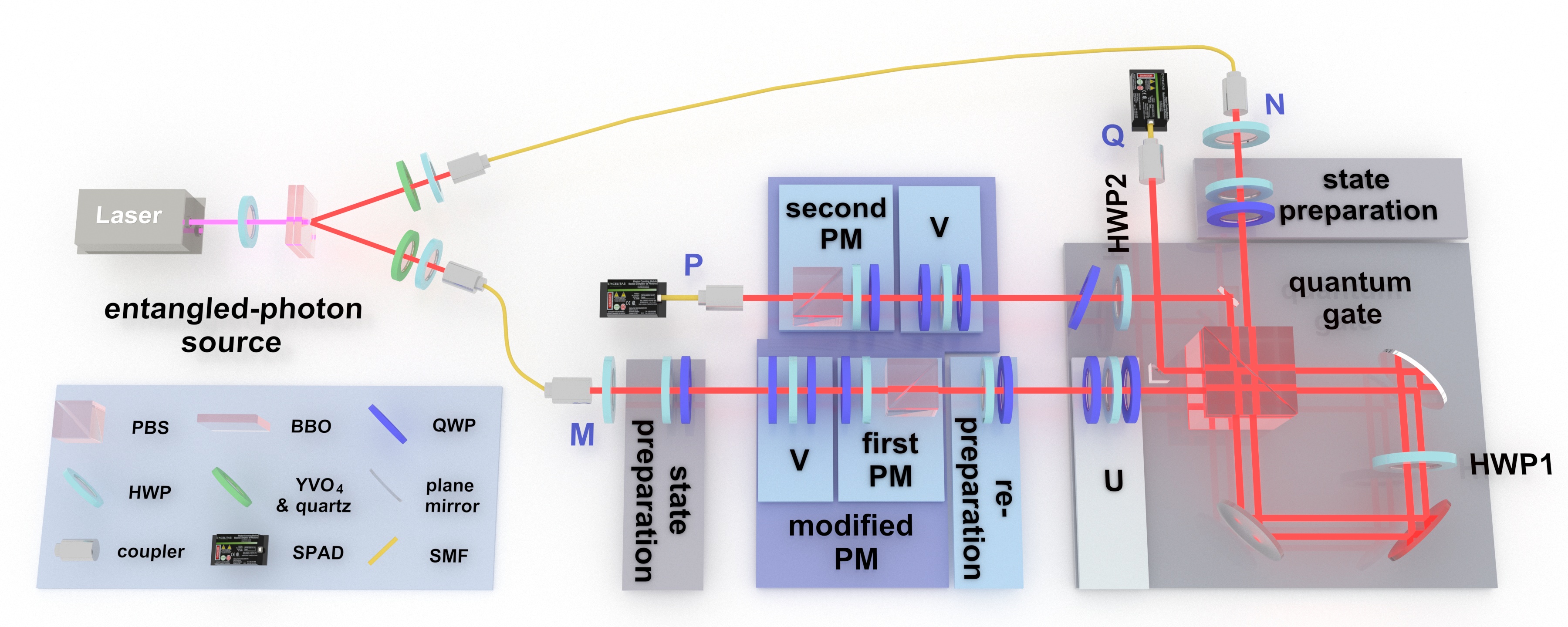}
    \caption{\textbf{Experimental setup.} The checkpoints M, N, P, and Q match the notations in Fig.\,\ref{fig_scheme}. 
    The monochromatic panels were inside the quantum comb and the blue panels were accessible to the causal discovery algorithm. 
    Entangled photons were generated by spontaneous parametric down-conversion. 
    The initial state $\rho$, the unitary channel $U$, and the modifications of Pauli measurement (PM) basis $V$ were realized by groups of wave plates, while the PMs themselves exploited an additional polarization beam splitter (PBS). The \textsf{SWAP} gate controlled by HWP1 and HWP2 can switch between DC and CC and the tilted QWP was for correcting the intra-gate polarization dispersion. HWP: half-wave plate; QWP: quarter-wave plate; SMF: single-mode fiber; SPAD: single-photon avalanche detectors.}
    \label{fig:setup}
 \end{figure*}
 
\textit{Algorithmic causal identification.}---As the two tetrahedra in Fig.\,\ref{fig_scheme}d are non-identical, some vectors $\mathbf{P}$ observed in the disjoint areas will imply causation\,\cite{Fitzsimons15}. However, the DC- and CC-tetrahedra also overlap in an octahedron. For those $\mathbf{P}$ falling there, both DC and CC causal explanations are possible, and, the Pauli correlations alone are insufficient to reveal causality. Here we show how to remove this ambiguity by using only a few additional measurements.
The starting point of our causal identification algorithm is that the cross-sections of the DC- and CC-tetrahedra at $C_{33}=1$ are two perpendicular lines and are distinguishable almost everywhere. Motivated by this observation, we seek to find a new set of observables $\sigma_k \to V\sigma_k V^\dagger, k\in\{1,2,3\}$ determined by the unitary operator $V$, so the correlation vectors for the new observables become $\mathbf{P}_{V}(U)=\mathbf{P}(V^\dagger UV)$ and $\mathbf{P}_{V}(\rho)=\mathbf{P}((V\otimes V)^\dagger \rho (V\otimes V))$. The goal of choosing $V$ is to make the third entry of $\mathbf{P}_{V}(U)$ equal to 1. To put it simply, the choice of $V$ should cause the channel to preserve the eigenstates of $V\sigma_3 V^\dagger$. 

The existence of such an operator $V$ is intuitive: in the Bloch sphere representation, a qubit unitary channel behaves as a rotation, and the states on the rotational axis are invariant. Therefore, an appropriate operator $V$ needs to change the orientation of the coordinate system so its zenith would align with the axis of rotation. Its explicit settings can be obtained from the Pauli correlations $\mathbf{P}$, which needs to be measured in the first place. 
The key observation is that, under the new measurement settings, any two-qubit DC correlation $\mathbf{P}_V(U)$ will satisfy $C_{33}^V(U)=1$ and $C_{11}^V(U)=C_{22}^V(U)$, where the sub/superscripts mean the measured observables are the Pauli operators modified by $V$; 
in contrast, only a zero-measure set of $\mathbf{P}(\rho)$ on an ``exceptional'' plane, $\sum_{k=1}^3 C_{kk}=1$ makes $C_{33}^V(\rho)=1$. Therefore, for those $\mathbf{P}$ not on the exceptional plane, the condition $C_{33}^V=1$ alone can serve as an indicator for the DC causal structure. We defer the numerical setting of $V$, the proof of the above observation, and the pseudocode for the entire algorithm to Supplemental Material\,\cite{SM}. 
\phantom{\cite{CHOI1975285, order2022}}

In practice, both the causal mechanism to be identified and the measurements themselves will be subject to noise. When applying the causal identification algorithm, we have to relax the condition for DC as $1-C_{33}^V<\varepsilon$ to allow some tolerance, where $\varepsilon>0$ is a cutoff value. The nonzero tolerance causes the indistinguishable CC set to have a nonzero measure. To handle this issue, whenever the initial correlation $\mathbf{P}$ is close to the exceptional plane ($1-\sum_{k=1}^3 C_{kk}<\delta$ where $\delta$ is a threshold value), we will modify the first measurement with $V$ and the second measurement with $V^\prime = \sigma_1 V$, take the resulting correlation as the new initial correlation and run the algorithm once again. Crucially, as the additional $\sigma_1$ enforces $C_{33}^V=-1$, the new initial correlation will be opposite to the exceptional plane on the octahedron. The final correlation induced by any DC, $\mathbf{P}_{V^\prime}(U)$, will thus always end close to $\mathbf{P}(\sigma_3)=\{-1, -1, 1\}$ far from the CC-tetrahedron, and reflect causality with strong resilience of noise. 
 
 \begin{figure*}[t]
    \centering
    \includegraphics[width=.99\textwidth]{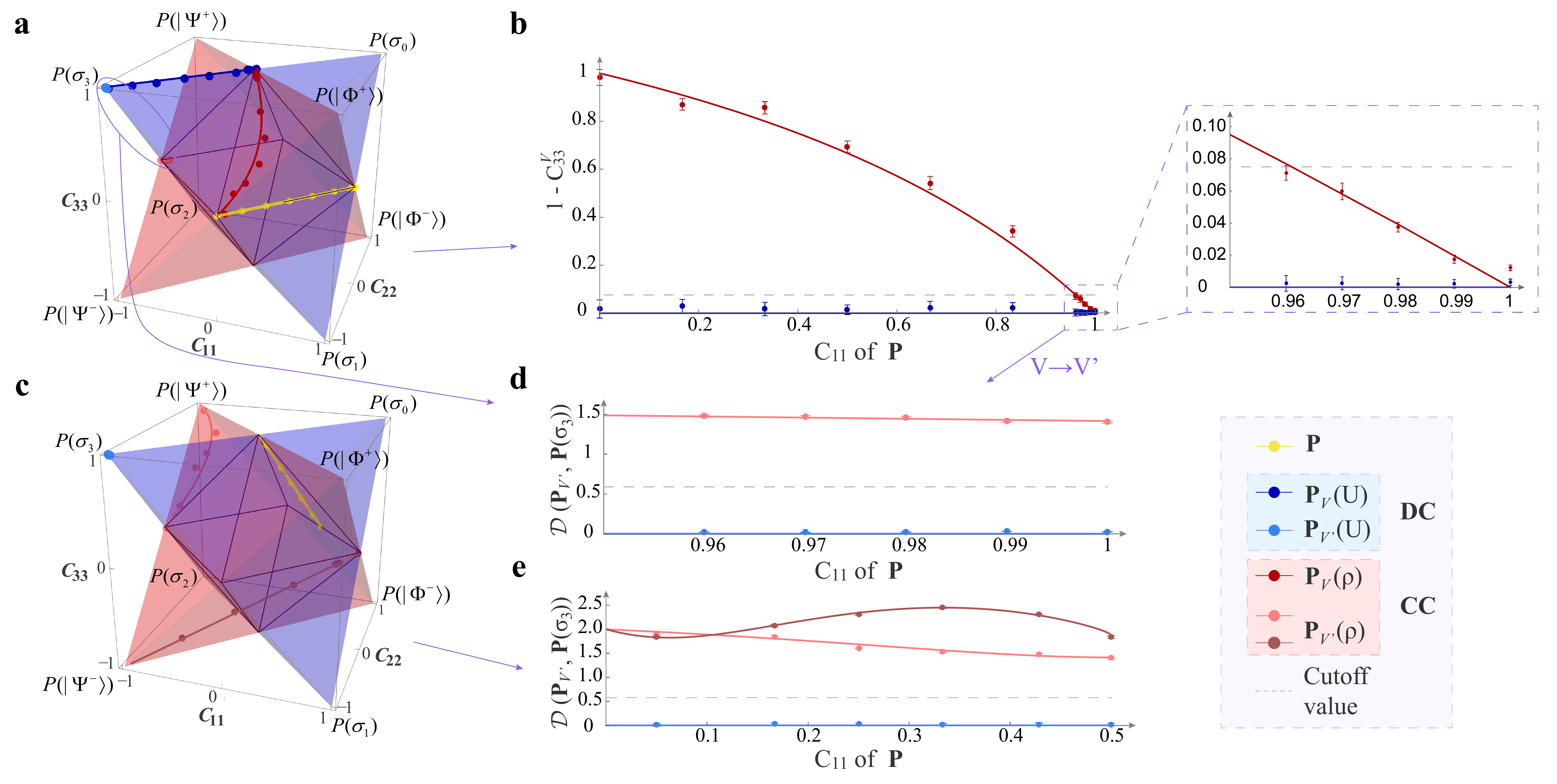}
    \caption{\textbf{Experimental results.} 
    \textbf{a}: Visualization of $\mathbf{P}_V$ when $\mathbf{P}$ moves on one edge of the causally indistinguishable octahedron. The $\mathbf{P}_{V^\prime}$ are evaluated only when the $\mathbf{P}$ are close to the exceptional plane. \textbf{b}: Predicted (line) and measured (data points) DC-criterion correspond to $\mathbf{P}_V$. The inset enlarges the indistinguishable part. \textbf{c}: Visualization of $\mathbf{P}_{V^\prime}$ when $\mathbf{P}$ moves on the exceptional plane. \textbf{d, e}: Predicted and measured distance criterion for DC correspond to the points $\mathbf{P}_{V^\prime}$ in \textbf{a} and \textbf{c}, respectively. The pink and brown lines show the effect of different phases of Bell states in state preparation. Error bars are calculated via the Monte Carlo method, and except for the 1$\sigma$ in the inset of \textbf{b}, all the other error bars are 3$\sigma$.
    }
    \label{fig:data}
\end{figure*}



\textit{Experiment.}---We implemented the quantum causal identification algorithm in an optical platform. We encoded two qubits on the polarization state of a photon pair generated from spontaneous parametric down-conversion, with the horizontal ($\ket{H}$) and vertical ($\ket{V}$) polarization states being the computational basis.
The layout of our experimental setup is depicted in Fig.\,\ref{fig:setup}. It closely resembled the quantum comb in Fig.\,\ref{fig_scheme}c; we have labeled four control points, M, N, P, and Q, in both of the figures to clarify the correspondence. Using the photon source reported in Reference\,\cite{czhang15}, we obtained the maximally-entangled state $(\ket{HH}+\ket{VV})/\sqrt{2}$ with fidelity over 0.97. 
Subsequently, we utilized two groups of half-wave plate (HWP) and quarter-wave plate (QWP) to prepare the desired initial states, and an HWP between two QWPs to implement the unitary channel operation\,\cite{Englert01}, so both of the operations would induce the same Pauli operations $\mathbf{P}$. 

A Sagnac-type ring interferometer at the center of the setup was exploited to switch the causal structure. 
Specifically, the polarizing beam splitter (PBS) at the entrance of the interferometer transmits (reflects) the photons with $\ket{H}$ ($\ket{V}$)-polarization. When the orientation of HWP1 and HWP2 were set at $0^{\circ}$($45^{\circ}$), the photon entered the setup from M will exit at P(Q) while keeping its polarization, so the ring would act as an identity (\textsf{SWAP}) gate and cause the photons at $X$ and $Y$ to be linked by DC (CC). 
Before and after the ring, we inserted a polarization measurement module containing a QWP, an HWP, and a PBS to implement a single-qubit projection onto the eigenstates of the Pauli operators. With two-photon coincidence counting, we can extract all the required conditional probabilities and determine the two-point correlations. The wave plates after the PBS in the first measurement reprepared the Pauli eigenstates. Due to the high degree of two-photon entanglement, the marginal state of one of the photons was maximally mixed, so the procedure did not introduce signaling and the causal discovery was strictly observational. Another group of wave plates introduced the $V$- and $V^\prime$-operations to modify the measurement directions. 

We tested two families of temporally-ordered quantum systems, whose causality is not unveiled by their Pauli correlations, to demonstrate the quantum causal discovery algorithm. The first family satisfies $C_{22}-C_{11}=1, C_{33}=0$; geometrically, the correlation is located on an edge of the indistinguishable octahedron where the DC- and CC-tetrahedra overlap. 
Although the two causal mechanisms can induce the same Pauli correlations, their induced correlations using the algorithmically-determined operators are drastically different: in Fig.\,\ref{fig:data}a, we plot the trajectories of the Pauli correlation $\mathbf{P}$ and the correlations under the modified measurement settings, $\mathbf{P}_V(U)$ (blue) and $\mathbf{P}_V(\rho)$ (red). We find that whenever the causal structure of $\mathbf{P}$ is DC, $\mathbf{P}_V(U)$ will be close to the $C_{33}^V(U)=1$ plane. The observation is quantitatively supported using the DC-criterion $1-C_{33}^V<\varepsilon$ as shown in Fig.\,\ref{fig:data}b, where we set $\delta=0.15$ and the cutoff value $\varepsilon=0.075$ successfully delineated all DC and CC far away from the exceptional plane. In Supplemental Material\,\cite{SM}, we also study the case of using mixed states as the source of CC to enrich the validity of this method and discuss its possible quantum advantage\,\cite{SM,mengyu,Chiribella13,Goswami18,Caslav14}. 


Finally, we corroborate that the algorithm can robustly handle the envisaged exception for Pauli correlations close to $\sum_{k=1}^3 C_{kk}=1$. As shown in the inset of Fig.\,\ref{fig:data}b, the DC-criterion using $P_V$ as input fails to identify the CC; however, the issue is fully resolved by running a second round of the algorithm. The resulted new correlations (cf. Fig.\,\ref{fig:data}a), $\mathbf{P}_{V^\prime}(U)$ (cyan) and $\mathbf{P}_{V^\prime}(\rho)$ (pink) definitely fall into the distinguishable regions. We further studied the case of another family of Pauli correlations moving on the exceptional plane, with the result depicted in Fig.\,\ref{fig:data}c. Clearly, the separation of $\mathbf{P}_{V^\prime}(U)$ and $\mathbf{P}_{V^\prime}(\rho)$ is robust and insensitive of the exact form of the channel/state.
To quantify the robustness of the algorithm we resort to another distance criterion for DC, ${\cal D}(\mathbf{P}_{V^\prime},\mathbf{P}(\sigma_3))<\varepsilon^\prime$. Here, the calligraphic $\cal D$ indicates the Euclidean distance of the two correlation vectors, and we conveniently select $\varepsilon^\prime=1/\sqrt{3}$. 
The results corresponding to the points running into and on the exceptional plane are given in Fig.\,\ref{fig:data}d and e. For all data points, the causal mechanisms were correctly identified beyond 70 standard deviations from the cutoff value. The results clearly demonstrated the stability of the quantum causal identification algorithm.

\textit{Conclusion.}---We have experimentally identified the underlying causal structures of two-point quantum correlations on a photonic platform. This algorithm is based on quantum correlation and unitary operations under the observational scheme, with no need for interventions compared to its classical counterpart and costs fewer resources. Quantum causal inference, as a relatively young field, is not only a quantum version of classical cause and effect that shows the significance of discovering the intrinsic causation between quantum variables; but also an interdisciplinary subject that provides a new perspective to investigate, characterize, and classify quantum correlations. We believe that our in-depth causal study of two-point quantum correlation will contribute to further developing quantum causal inferences and even constructing causal networks between quantum resources. 

\begin{acknowledgments}
\textit{Acknowledgments.}---This work was supported by the Innovation Program for Quantum Science and Technology (No. 2021ZD0301200), the National Natural Science Foundation of China (Grants 12174370, 12174376, 11821404, and 11904356), the Youth Innovation Promotion Association of Chinese Academy of Sciences (Grant No. 2017492), the Open Research Projects of Zhejiang Lab (No. 2021MB0AB02), the Fok Ying-Tong Education Foundation (No. 171007), and the Fundamental Research Funds for the Central Universities (No. WK2030000008). 
\end{acknowledgments}

\input{references.bbl}

\end{document}

%% file: references.bbl
%

%% file: CA-V3.1-2023.bbl
\begin{thebibliography}{30}%
\makeatletter
\providecommand \@ifxundefined [1]{%
 \@ifx{#1\undefined}
}%
\providecommand \@ifnum [1]{%
 \ifnum #1\expandafter \@firstoftwo
 \else \expandafter \@secondoftwo
 \fi
}%
\providecommand \@ifx [1]{%
 \ifx #1\expandafter \@firstoftwo
 \else \expandafter \@secondoftwo
 \fi
}%
\providecommand \natexlab [1]{#1}%
\providecommand \enquote  [1]{``#1''}%
\providecommand \bibnamefont  [1]{#1}%
\providecommand \bibfnamefont [1]{#1}%
\providecommand \citenamefont [1]{#1}%
\providecommand \href@noop [0]{\@secondoftwo}%
\providecommand \href [0]{\begingroup \@sanitize@url \@href}%
\providecommand \@href[1]{\@@startlink{#1}\@@href}%
\providecommand \@@href[1]{\endgroup#1\@@endlink}%
\providecommand \@sanitize@url [0]{\catcode `\\12\catcode `\$12\catcode
  `\&12\catcode `\#12\catcode `\^12\catcode `\_12\catcode `\%12\relax}%
\providecommand \@@startlink[1]{}%
\providecommand \@@endlink[0]{}%
\providecommand \url  [0]{\begingroup\@sanitize@url \@url }%
\providecommand \@url [1]{\endgroup\@href {#1}{\urlprefix }}%
\providecommand \urlprefix  [0]{URL }%
\providecommand \Eprint [0]{\href }%
\providecommand \doibase [0]{https://doi.org/}%
\providecommand \selectlanguage [0]{\@gobble}%
\providecommand \bibinfo  [0]{\@secondoftwo}%
\providecommand \bibfield  [0]{\@secondoftwo}%
\providecommand \translation [1]{[#1]}%
\providecommand \BibitemOpen [0]{}%
\providecommand \bibitemStop [0]{}%
\providecommand \bibitemNoStop [0]{.\EOS\space}%
\providecommand \EOS [0]{\spacefactor3000\relax}%
\providecommand \BibitemShut  [1]{\csname bibitem#1\endcsname}%
\let\auto@bib@innerbib\@empty
\bibitem [{\citenamefont {DeGrave}\ \emph {et~al.}(2021)\citenamefont
  {DeGrave}, \citenamefont {Janizek},\ and\ \citenamefont {Lee}}]{DeGrave21}%
  \BibitemOpen
  \bibfield  {author} {\bibinfo {author} {\bibfnamefont {A.~J.}\ \bibnamefont
  {DeGrave}}, \bibinfo {author} {\bibfnamefont {J.~D.}\ \bibnamefont
  {Janizek}},\ and\ \bibinfo {author} {\bibfnamefont {S.-I.}\ \bibnamefont
  {Lee}},\ }\bibfield  {title} {\bibinfo {title} {Ai for radiographic covid-19
  detection selects shortcuts over signal},\ }\href@noop {} {\bibfield
  {journal} {\bibinfo  {journal} {Nature Machine Intelligence}\ }\textbf
  {\bibinfo {volume} {3}},\ \bibinfo {pages} {610} (\bibinfo {year}
  {2021})}\BibitemShut {NoStop}%
\bibitem [{\citenamefont {Savage}(2023)}]{Savage23}%
  \BibitemOpen
  \bibfield  {author} {\bibinfo {author} {\bibfnamefont {N.}~\bibnamefont
  {Savage}},\ }\bibfield  {title} {\bibinfo {title} {Why artificial
  intelligence needs to understand consequences.},\ }\href
  {https://doi.org/10.1038/d41586-023-00577-1} {\bibfield  {journal} {\bibinfo
  {journal} {Nature}\ } (\bibinfo {year} {2023})}\BibitemShut {NoStop}%
\bibitem [{\citenamefont {Cui}\ and\ \citenamefont
  {Athey}(2022)}]{cui2022stable}%
  \BibitemOpen
  \bibfield  {author} {\bibinfo {author} {\bibfnamefont {P.}~\bibnamefont
  {Cui}}\ and\ \bibinfo {author} {\bibfnamefont {S.}~\bibnamefont {Athey}},\
  }\bibfield  {title} {\bibinfo {title} {Stable learning establishes some
  common ground between causal inference and machine learning},\ }\href@noop {}
  {\bibfield  {journal} {\bibinfo  {journal} {Nature Machine Intelligence}\
  }\textbf {\bibinfo {volume} {4}},\ \bibinfo {pages} {110} (\bibinfo {year}
  {2022})}\BibitemShut {NoStop}%
\bibitem [{\citenamefont {Reichenbach}\ and\ \citenamefont
  {Reichenbach}(1991)}]{Reichenbach}%
  \BibitemOpen
  \bibfield  {author} {\bibinfo {author} {\bibfnamefont {H.}~\bibnamefont
  {Reichenbach}}\ and\ \bibinfo {author} {\bibfnamefont {M.}~\bibnamefont
  {Reichenbach}},\ }\href@noop {} {\emph {\bibinfo {title} {The Direction of
  Time}}}\ (\bibinfo  {publisher} {University of California Press},\ \bibinfo
  {year} {1991})\BibitemShut {NoStop}%
\bibitem [{\citenamefont {Bell}(1964)}]{Bell64}%
  \BibitemOpen
  \bibfield  {author} {\bibinfo {author} {\bibfnamefont {J.~S.}\ \bibnamefont
  {Bell}},\ }\bibfield  {title} {\bibinfo {title} {On the einstein podolsky
  rosen paradox},\ }\href {https://doi.org/10.1103/PhysicsPhysiqueFizika.1.195}
  {\bibfield  {journal} {\bibinfo  {journal} {Phys. Phys. Fiz.}\ }\textbf
  {\bibinfo {volume} {1}},\ \bibinfo {pages} {195} (\bibinfo {year}
  {1964})}\BibitemShut {NoStop}%
\bibitem [{\citenamefont {Kochen}\ and\ \citenamefont {Specker}(1967)}]{KS67}%
  \BibitemOpen
  \bibfield  {author} {\bibinfo {author} {\bibfnamefont {S.}~\bibnamefont
  {Kochen}}\ and\ \bibinfo {author} {\bibfnamefont {E.~P.}\ \bibnamefont
  {Specker}},\ }\bibfield  {title} {\bibinfo {title} {The problem of hidden
  variables in quantum mechanics},\ }\href
  {https://www.jstor.org/stable/24902153} {\bibfield  {journal} {\bibinfo
  {journal} {J. Math. Mech.}\ }\textbf {\bibinfo {volume} {17}},\ \bibinfo
  {pages} {59} (\bibinfo {year} {1967})}\BibitemShut {NoStop}%
\bibitem [{\citenamefont {Leggett}\ and\ \citenamefont {Garg}(1985)}]{LG85}%
  \BibitemOpen
  \bibfield  {author} {\bibinfo {author} {\bibfnamefont {A.~J.}\ \bibnamefont
  {Leggett}}\ and\ \bibinfo {author} {\bibfnamefont {A.}~\bibnamefont {Garg}},\
  }\bibfield  {title} {\bibinfo {title} {Quantum mechanics versus macroscopic
  realism: {I}s the flux there when nobody looks?},\ }\href
  {https://doi.org/10.1103/PhysRevLett.54.857} {\bibfield  {journal} {\bibinfo
  {journal} {Phys. Rev. Lett.}\ }\textbf {\bibinfo {volume} {54}},\ \bibinfo
  {pages} {857} (\bibinfo {year} {1985})}\BibitemShut {NoStop}%
\bibitem [{\citenamefont {Cavalcanti}\ and\ \citenamefont
  {Lal}(2014)}]{Cavalcanti2014}%
  \BibitemOpen
  \bibfield  {author} {\bibinfo {author} {\bibfnamefont {E.~G.}\ \bibnamefont
  {Cavalcanti}}\ and\ \bibinfo {author} {\bibfnamefont {R.}~\bibnamefont
  {Lal}},\ }\bibfield  {title} {\bibinfo {title} {On modifications of
  reichenbach's principle of common cause in light of bell's theorem},\ }\href
  {https://doi.org/10.1088/1751-8113/47/42/424018} {\bibfield  {journal}
  {\bibinfo  {journal} {J. Phys. A: Math. Theor.}\ }\textbf {\bibinfo {volume}
  {47}},\ \bibinfo {pages} {424018} (\bibinfo {year} {2014})}\BibitemShut
  {NoStop}%
\bibitem [{\citenamefont {Henson}\ \emph {et~al.}(2014)\citenamefont {Henson},
  \citenamefont {Lal},\ and\ \citenamefont {Pusey}}]{Henson2014}%
  \BibitemOpen
  \bibfield  {author} {\bibinfo {author} {\bibfnamefont {J.}~\bibnamefont
  {Henson}}, \bibinfo {author} {\bibfnamefont {R.}~\bibnamefont {Lal}},\ and\
  \bibinfo {author} {\bibfnamefont {M.~F.}\ \bibnamefont {Pusey}},\ }\bibfield
  {title} {\bibinfo {title} {Theory-independent limits on correlations from
  generalized bayesian networks},\ }\href
  {https://doi.org/10.1088/1367-2630/16/11/113043} {\bibfield  {journal}
  {\bibinfo  {journal} {New J. Phys.}\ }\textbf {\bibinfo {volume} {16}},\
  \bibinfo {pages} {113043} (\bibinfo {year} {2014})}\BibitemShut {NoStop}%
\bibitem [{\citenamefont {Fitzsimons}\ \emph {et~al.}(2015)\citenamefont
  {Fitzsimons}, \citenamefont {Jones},\ and\ \citenamefont
  {Vedral}}]{Fitzsimons15}%
  \BibitemOpen
  \bibfield  {author} {\bibinfo {author} {\bibfnamefont {J.~F.}\ \bibnamefont
  {Fitzsimons}}, \bibinfo {author} {\bibfnamefont {J.~A.}\ \bibnamefont
  {Jones}},\ and\ \bibinfo {author} {\bibfnamefont {V.}~\bibnamefont
  {Vedral}},\ }\bibfield  {title} {\bibinfo {title} {Quantum correlations which
  imply causation},\ }\href {https://www.nature.com/articles/srep18281}
  {\bibfield  {journal} {\bibinfo  {journal} {Sci. Rep.}\ }\textbf {\bibinfo
  {volume} {5}},\ \bibinfo {pages} {18281} (\bibinfo {year}
  {2015})}\BibitemShut {NoStop}%
\bibitem [{\citenamefont {Ried}\ \emph {et~al.}(2015)\citenamefont {Ried},
  \citenamefont {Agnew}, \citenamefont {Vermeyden}, \citenamefont {Janzing},
  \citenamefont {Spekkens},\ and\ \citenamefont {Resch}}]{Ried15}%
  \BibitemOpen
  \bibfield  {author} {\bibinfo {author} {\bibfnamefont {K.}~\bibnamefont
  {Ried}}, \bibinfo {author} {\bibfnamefont {M.}~\bibnamefont {Agnew}},
  \bibinfo {author} {\bibfnamefont {L.}~\bibnamefont {Vermeyden}}, \bibinfo
  {author} {\bibfnamefont {D.}~\bibnamefont {Janzing}}, \bibinfo {author}
  {\bibfnamefont {R.~W.}\ \bibnamefont {Spekkens}},\ and\ \bibinfo {author}
  {\bibfnamefont {K.~J.}\ \bibnamefont {Resch}},\ }\bibfield  {title} {\bibinfo
  {title} {A quantum advantage for inferring causal structure},\ }\href
  {http://dx.doi.org/10.1038/nphys3266} {\bibfield  {journal} {\bibinfo
  {journal} {Nat. Phys.}\ }\textbf {\bibinfo {volume} {11}},\ \bibinfo {pages}
  {414} (\bibinfo {year} {2015})}\BibitemShut {NoStop}%
\bibitem [{\citenamefont {Chaves}\ \emph {et~al.}(2015)\citenamefont {Chaves},
  \citenamefont {Majenz},\ and\ \citenamefont {Gross}}]{Chaves2015}%
  \BibitemOpen
  \bibfield  {author} {\bibinfo {author} {\bibfnamefont {R.}~\bibnamefont
  {Chaves}}, \bibinfo {author} {\bibfnamefont {C.}~\bibnamefont {Majenz}},\
  and\ \bibinfo {author} {\bibfnamefont {D.}~\bibnamefont {Gross}},\ }\bibfield
   {title} {\bibinfo {title} {Information--theoretic implications of quantum
  causal structures},\ }\href {https://doi.org/10.1038/ncomms6766} {\bibfield
  {journal} {\bibinfo  {journal} {Nat. Commun.}\ }\textbf {\bibinfo {volume}
  {6}},\ \bibinfo {pages} {5766} (\bibinfo {year} {2015})}\BibitemShut
  {NoStop}%
\bibitem [{\citenamefont {Costa}\ and\ \citenamefont
  {Shrapnel}(2016)}]{Costa2016}%
  \BibitemOpen
  \bibfield  {author} {\bibinfo {author} {\bibfnamefont {F.}~\bibnamefont
  {Costa}}\ and\ \bibinfo {author} {\bibfnamefont {S.}~\bibnamefont
  {Shrapnel}},\ }\bibfield  {title} {\bibinfo {title} {Quantum causal
  modelling},\ }\href {https://doi.org/10.1088/1367-2630/18/6/063032}
  {\bibfield  {journal} {\bibinfo  {journal} {New J. Phys.}\ }\textbf {\bibinfo
  {volume} {18}},\ \bibinfo {pages} {063032} (\bibinfo {year}
  {2016})}\BibitemShut {NoStop}%
\bibitem [{\citenamefont {Allen}\ \emph {et~al.}(2017)\citenamefont {Allen},
  \citenamefont {Barrett}, \citenamefont {Horsman}, \citenamefont {Lee},\ and\
  \citenamefont {Spekkens}}]{Allen17}%
  \BibitemOpen
  \bibfield  {author} {\bibinfo {author} {\bibfnamefont {J.-M.~A.}\
  \bibnamefont {Allen}}, \bibinfo {author} {\bibfnamefont {J.}~\bibnamefont
  {Barrett}}, \bibinfo {author} {\bibfnamefont {D.~C.}\ \bibnamefont
  {Horsman}}, \bibinfo {author} {\bibfnamefont {C.~M.}\ \bibnamefont {Lee}},\
  and\ \bibinfo {author} {\bibfnamefont {R.~W.}\ \bibnamefont {Spekkens}},\
  }\bibfield  {title} {\bibinfo {title} {Quantum common cause and quantum
  causal models},\ }\href {https://doi.org/10.1103/PhysRevX.7.031021}
  {\bibfield  {journal} {\bibinfo  {journal} {Phys. Rev. X}\ }\textbf {\bibinfo
  {volume} {7}},\ \bibinfo {pages} {031021} (\bibinfo {year}
  {2017})}\BibitemShut {NoStop}%
\bibitem [{\citenamefont {Chaves}\ \emph {et~al.}(2018)\citenamefont {Chaves},
  \citenamefont {Carvacho}, \citenamefont {Agresti}, \citenamefont {Giulio},
  \citenamefont {Aolita}, \citenamefont {Giacomini},\ and\ \citenamefont
  {Sciarrino}}]{Chaves18}%
  \BibitemOpen
  \bibfield  {author} {\bibinfo {author} {\bibfnamefont {R.}~\bibnamefont
  {Chaves}}, \bibinfo {author} {\bibfnamefont {G.}~\bibnamefont {Carvacho}},
  \bibinfo {author} {\bibfnamefont {I.}~\bibnamefont {Agresti}}, \bibinfo
  {author} {\bibfnamefont {V.~D.}\ \bibnamefont {Giulio}}, \bibinfo {author}
  {\bibfnamefont {L.}~\bibnamefont {Aolita}}, \bibinfo {author} {\bibfnamefont
  {S.}~\bibnamefont {Giacomini}},\ and\ \bibinfo {author} {\bibfnamefont
  {F.}~\bibnamefont {Sciarrino}},\ }\bibfield  {title} {\bibinfo {title}
  {Quantum violation of an instrumental test},\ }\href
  {https://doi.org/10.1038/s41567-017-0008-5} {\bibfield  {journal} {\bibinfo
  {journal} {Nat. Phys.}\ }\textbf {\bibinfo {volume} {14}},\ \bibinfo {pages}
  {291} (\bibinfo {year} {2018})}\BibitemShut {NoStop}%
\bibitem [{\citenamefont {Hu}\ and\ \citenamefont {Hou}(2018)}]{mdhu18}%
  \BibitemOpen
  \bibfield  {author} {\bibinfo {author} {\bibfnamefont {M.}~\bibnamefont
  {Hu}}\ and\ \bibinfo {author} {\bibfnamefont {Y.}~\bibnamefont {Hou}},\
  }\bibfield  {title} {\bibinfo {title} {Discrimination between quantum common
  causes and quantum causality},\ }\href
  {https://doi.org/10.1103/PhysRevA.97.062125} {\bibfield  {journal} {\bibinfo
  {journal} {Phys. Rev. A}\ }\textbf {\bibinfo {volume} {97}},\ \bibinfo
  {pages} {062125} (\bibinfo {year} {2018})}\BibitemShut {NoStop}%
\bibitem [{\citenamefont {Barrett}\ \emph {et~al.}(2019)\citenamefont
  {Barrett}, \citenamefont {Lorenz},\ and\ \citenamefont
  {Oreshkov}}]{barrett2019}%
  \BibitemOpen
  \bibfield  {author} {\bibinfo {author} {\bibfnamefont {J.}~\bibnamefont
  {Barrett}}, \bibinfo {author} {\bibfnamefont {R.}~\bibnamefont {Lorenz}},\
  and\ \bibinfo {author} {\bibfnamefont {O.}~\bibnamefont {Oreshkov}},\
  }\bibfield  {title} {\bibinfo {title} {Quantum causal models},\ }\Eprint
  {https://arxiv.org/abs/1906.10726} {arXiv:1906.10726 [quant-ph]}  (\bibinfo
  {year} {2019})\BibitemShut {NoStop}%
\bibitem [{\citenamefont {Zhang}\ \emph {et~al.}(2020)\citenamefont {Zhang},
  \citenamefont {Hou},\ and\ \citenamefont {Song}}]{cgzhang20}%
  \BibitemOpen
  \bibfield  {author} {\bibinfo {author} {\bibfnamefont {C.}~\bibnamefont
  {Zhang}}, \bibinfo {author} {\bibfnamefont {Y.}~\bibnamefont {Hou}},\ and\
  \bibinfo {author} {\bibfnamefont {D.}~\bibnamefont {Song}},\ }\bibfield
  {title} {\bibinfo {title} {Quantum observation scheme universally identifying
  causalities from correlations},\ }\href
  {https://doi.org/10.1103/PhysRevA.101.062103} {\bibfield  {journal} {\bibinfo
   {journal} {Phys. Rev. A}\ }\textbf {\bibinfo {volume} {101}},\ \bibinfo
  {pages} {062103} (\bibinfo {year} {2020})}\BibitemShut {NoStop}%
\bibitem [{\citenamefont {Chiribella}\ \emph {et~al.}(2007)\citenamefont
  {Chiribella}, \citenamefont {D'Ariano},\ and\ \citenamefont
  {Perinotti}}]{Chiribella07}%
  \BibitemOpen
  \bibfield  {author} {\bibinfo {author} {\bibfnamefont {G.}~\bibnamefont
  {Chiribella}}, \bibinfo {author} {\bibfnamefont {G.~M.}\ \bibnamefont
  {D'Ariano}},\ and\ \bibinfo {author} {\bibfnamefont {P.}~\bibnamefont
  {Perinotti}},\ }\bibfield  {title} {\bibinfo {title} {Quantum circuit
  architecture},\ }\href
  {https://journals.aps.org/prl/abstract/10.1103/PhysRevLett.101.060401}
  {\bibfield  {journal} {\bibinfo  {journal} {Phys. Rev. Lett.}\ }\textbf
  {\bibinfo {volume} {101}},\ \bibinfo {pages} {060401} (\bibinfo {year}
  {2007})}\BibitemShut {NoStop}%
\bibitem [{\citenamefont {Chiribella}\ and\ \citenamefont
  {Swati}(2021)}]{Chiribella21}%
  \BibitemOpen
  \bibfield  {author} {\bibinfo {author} {\bibfnamefont {G.}~\bibnamefont
  {Chiribella}}\ and\ \bibinfo {author} {\bibnamefont {Swati}},\ }\bibfield
  {title} {\bibinfo {title} {Fast tests for probing the causal structure of
  quantum processes},\ }in\ \href@noop {} {\emph {\bibinfo {booktitle} {Quantum
  Theory and Symmetries}}}\ (\bibinfo  {publisher} {Springer International
  Publishing},\ \bibinfo {address} {Cham},\ \bibinfo {year} {2021})\ pp.\
  \bibinfo {pages} {617--632}\BibitemShut {NoStop}%
\bibitem [{\citenamefont {L\"{u}ders}(1950)}]{Luders}%
  \BibitemOpen
  \bibfield  {author} {\bibinfo {author} {\bibfnamefont {G.}~\bibnamefont
  {L\"{u}ders}},\ }\bibfield  {title} {\bibinfo {title} {\"{U}ber die
  zustands\"anderung durch den messproze\ss},\ }\href
  {https://doi.org/10.1002/andp.19504430510} {\bibfield  {journal} {\bibinfo
  {journal} {Ann. Phys. (Berlin)}\ }\textbf {\bibinfo {volume} {443}},\
  \bibinfo {pages} {322} (\bibinfo {year} {1950})}\BibitemShut {NoStop}%
\bibitem [{SM()}]{SM}%
  \BibitemOpen
  \bibinfo {title} {See supplemental material for theoretical details and
  additional experimental results, which also includes references
  \cite{CHOI1975285} and \cite{order2022}}\BibitemShut {NoStop}%
\bibitem [{\citenamefont {Choi}(1975)}]{CHOI1975285}%
  \BibitemOpen
\bibfield  {title} {  }\bibfield  {author} {\bibinfo {author} {\bibfnamefont
  {M.-D.}\ \bibnamefont {Choi}},\ }\bibfield  {title} {\bibinfo {title}
  {Completely positive linear maps on complex matrices},\ }\href
  {https://doi.org/10.1016/0024-3795(75)90075-0} {\bibfield  {journal}
  {\bibinfo  {journal} {Linear Algebra Appl.}\ }\textbf {\bibinfo {volume}
  {10}},\ \bibinfo {pages} {285} (\bibinfo {year} {1975})}\BibitemShut
  {NoStop}%
\bibitem [{\citenamefont {Dourdent}\ \emph {et~al.}(2022)\citenamefont
  {Dourdent}, \citenamefont {Abbott}, \citenamefont {Brunner}, \citenamefont
  {\ifmmode \check{S}\else \v{S}\fi{}upi\ifmmode~\acute{c}\else \'{c}\fi{}},\
  and\ \citenamefont {Branciard}}]{order2022}%
  \BibitemOpen
  \bibfield  {author} {\bibinfo {author} {\bibfnamefont {H.}~\bibnamefont
  {Dourdent}}, \bibinfo {author} {\bibfnamefont {A.~A.}\ \bibnamefont
  {Abbott}}, \bibinfo {author} {\bibfnamefont {N.}~\bibnamefont {Brunner}},
  \bibinfo {author} {\bibfnamefont {I.}~\bibnamefont {\ifmmode \check{S}\else
  \v{S}\fi{}upi\ifmmode~\acute{c}\else \'{c}\fi{}}},\ and\ \bibinfo {author}
  {\bibfnamefont {C.}~\bibnamefont {Branciard}},\ }\bibfield  {title} {\bibinfo
  {title} {Semi-device-independent certification of causal nonseparability with
  trusted quantum inputs},\ }\href
  {https://doi.org/10.1103/PhysRevLett.129.090402} {\bibfield  {journal}
  {\bibinfo  {journal} {Phys. Rev. Lett.}\ }\textbf {\bibinfo {volume} {129}},\
  \bibinfo {pages} {090402} (\bibinfo {year} {2022})}\BibitemShut {NoStop}%
\bibitem [{\citenamefont {Zhang}\ \emph {et~al.}(2015)\citenamefont {Zhang},
  \citenamefont {Huang}, \citenamefont {Wang}, \citenamefont {Liu},
  \citenamefont {Li},\ and\ \citenamefont {Guo}}]{czhang15}%
  \BibitemOpen
  \bibfield  {author} {\bibinfo {author} {\bibfnamefont {C.}~\bibnamefont
  {Zhang}}, \bibinfo {author} {\bibfnamefont {Y.-F.}\ \bibnamefont {Huang}},
  \bibinfo {author} {\bibfnamefont {Z.}~\bibnamefont {Wang}}, \bibinfo {author}
  {\bibfnamefont {B.-H.}\ \bibnamefont {Liu}}, \bibinfo {author} {\bibfnamefont
  {C.-F.}\ \bibnamefont {Li}},\ and\ \bibinfo {author} {\bibfnamefont {G.-C.}\
  \bibnamefont {Guo}},\ }\bibfield  {title} {\bibinfo {title} {Experimental
  {G}reenberger-{H}orne-{Z}eilinger-type six-photon quantum nonlocality},\
  }\href {https://doi.org/10.1103/PhysRevLett.115.260402} {\bibfield  {journal}
  {\bibinfo  {journal} {Phys. Rev. Lett.}\ }\textbf {\bibinfo {volume} {115}},\
  \bibinfo {pages} {260402} (\bibinfo {year} {2015})}\BibitemShut {NoStop}%
\bibitem [{\citenamefont {Englert}\ \emph {et~al.}(2001)\citenamefont
  {Englert}, \citenamefont {Kurtsiefer},\ and\ \citenamefont
  {Weinfurter}}]{Englert01}%
  \BibitemOpen
  \bibfield  {author} {\bibinfo {author} {\bibfnamefont {B.-G.}\ \bibnamefont
  {Englert}}, \bibinfo {author} {\bibfnamefont {C.}~\bibnamefont
  {Kurtsiefer}},\ and\ \bibinfo {author} {\bibfnamefont {H.}~\bibnamefont
  {Weinfurter}},\ }\bibfield  {title} {\bibinfo {title} {Universal unitary gate
  for single-photon two-qubit states},\ }\href
  {https://doi.org/10.1103/PhysRevA.63.032303} {\bibfield  {journal} {\bibinfo
  {journal} {Phys. Rev. A}\ }\textbf {\bibinfo {volume} {63}},\ \bibinfo
  {pages} {032303} (\bibinfo {year} {2001})}\BibitemShut {NoStop}%
\bibitem [{\citenamefont {Meng}\ \emph {et~al.}(2021)\citenamefont {Meng},
  \citenamefont {Liu}, \citenamefont {Zhao}, \citenamefont {Yin}, \citenamefont
  {Wang}, \citenamefont {Liu}, \citenamefont {Li}, \citenamefont {Yang},
  \citenamefont {Wang}, \citenamefont {Xu}, \citenamefont {Yu}, \citenamefont
  {Tang}, \citenamefont {Li},\ and\ \citenamefont {Guo}}]{mengyu}%
  \BibitemOpen
  \bibfield  {author} {\bibinfo {author} {\bibfnamefont {Y.}~\bibnamefont
  {Meng}}, \bibinfo {author} {\bibfnamefont {Z.-H.}\ \bibnamefont {Liu}},
  \bibinfo {author} {\bibfnamefont {Z.-K.}\ \bibnamefont {Zhao}}, \bibinfo
  {author} {\bibfnamefont {P.}~\bibnamefont {Yin}}, \bibinfo {author}
  {\bibfnamefont {Y.-T.}\ \bibnamefont {Wang}}, \bibinfo {author}
  {\bibfnamefont {W.}~\bibnamefont {Liu}}, \bibinfo {author} {\bibfnamefont
  {Z.-P.}\ \bibnamefont {Li}}, \bibinfo {author} {\bibfnamefont {Y.-Z.}\
  \bibnamefont {Yang}}, \bibinfo {author} {\bibfnamefont {Z.-A.}\ \bibnamefont
  {Wang}}, \bibinfo {author} {\bibfnamefont {J.-S.}\ \bibnamefont {Xu}},
  \bibinfo {author} {\bibfnamefont {S.}~\bibnamefont {Yu}}, \bibinfo {author}
  {\bibfnamefont {J.-S.}\ \bibnamefont {Tang}}, \bibinfo {author}
  {\bibfnamefont {C.-F.}\ \bibnamefont {Li}},\ and\ \bibinfo {author}
  {\bibfnamefont {G.-C.}\ \bibnamefont {Guo}},\ }\bibfield  {title} {\bibinfo
  {title} {Probing asymmetry in spatial-temporal correlations in quantum causal
  inference},\ }\href {https://www.researchsquare.com/article/rs-311195/v1}
  {\bibfield  {journal} {\bibinfo  {journal} {Research Square}\ } (\bibinfo
  {year} {2021})}\BibitemShut {NoStop}%
\bibitem [{\citenamefont {Chiribella}\ \emph {et~al.}(2013)\citenamefont
  {Chiribella}, \citenamefont {Mauro~D'Ariano}, \citenamefont {Perinotti},\
  and\ \citenamefont {Valiron}}]{Chiribella13}%
  \BibitemOpen
  \bibfield  {author} {\bibinfo {author} {\bibfnamefont {G.}~\bibnamefont
  {Chiribella}}, \bibinfo {author} {\bibfnamefont {G.}~\bibnamefont
  {Mauro~D'Ariano}}, \bibinfo {author} {\bibfnamefont {P.}~\bibnamefont
  {Perinotti}},\ and\ \bibinfo {author} {\bibfnamefont {B.}~\bibnamefont
  {Valiron}},\ }\bibfield  {title} {\bibinfo {title} {Quantum computations
  without definite causal structure},\ }\href
  {https://doi.org/10.1103/PhysRevA.88.022318} {\bibfield  {journal} {\bibinfo
  {journal} {Phys. Rev. A}\ }\textbf {\bibinfo {volume} {88}},\ \bibinfo
  {pages} {022318} (\bibinfo {year} {2013})}\BibitemShut {NoStop}%
\bibitem [{\citenamefont {Goswami}\ \emph {et~al.}(2018)\citenamefont
  {Goswami}, \citenamefont {Giarmatzi}, \citenamefont {Kewming}, \citenamefont
  {Costa}, \citenamefont {Branciard}, \citenamefont {Romero},\ and\
  \citenamefont {White}}]{Goswami18}%
  \BibitemOpen
  \bibfield  {author} {\bibinfo {author} {\bibfnamefont {K.}~\bibnamefont
  {Goswami}}, \bibinfo {author} {\bibfnamefont {C.}~\bibnamefont {Giarmatzi}},
  \bibinfo {author} {\bibfnamefont {M.}~\bibnamefont {Kewming}}, \bibinfo
  {author} {\bibfnamefont {F.}~\bibnamefont {Costa}}, \bibinfo {author}
  {\bibfnamefont {C.}~\bibnamefont {Branciard}}, \bibinfo {author}
  {\bibfnamefont {J.}~\bibnamefont {Romero}},\ and\ \bibinfo {author}
  {\bibfnamefont {A.}~\bibnamefont {White}},\ }\bibfield  {title} {\bibinfo
  {title} {Indefinite causal order in a quantum switch},\ }\href
  {https://doi.org/10.1103/PhysRevA.88.022318} {\bibfield  {journal} {\bibinfo
  {journal} {Phys. Rev. Lett.}\ }\textbf {\bibinfo {volume} {121}},\ \bibinfo
  {pages} {090503} (\bibinfo {year} {2018})}\BibitemShut {NoStop}%
\bibitem [{\citenamefont {\v{C} Brukner}(2014)}]{Caslav14}%
  \BibitemOpen
  \bibfield  {author} {\bibinfo {author} {\bibnamefont {\v{C} Brukner}},\
  }\bibfield  {title} {\bibinfo {title} {Quantum causality},\ }\href
  {https://doi.org/10.1038/nphys2930} {\bibfield  {journal} {\bibinfo
  {journal} {Nat. Phys.}\ }\textbf {\bibinfo {volume} {10}},\ \bibinfo {pages}
  {259} (\bibinfo {year} {2014})}\BibitemShut {NoStop}%
\end{thebibliography}
